\documentstyle[preprint,aps]{revtex}
\begin{document}
\preprint{DOE/ER/40427-16-N94}
\draft
\title{
COLLAPSE OF FLUX TUBES}
\author{L. Wilets and R. D. Puff}
\address{Department of Physics, University of Washington,
Seattle, WA 98195, USA}
\date{\today}
\maketitle
\begin{abstract}

      The dynamics of an idealized, infinite, MIT-type flux tube is
followed in time as the interior evolves from a pure gluon field to a
$\overline q \ q$ plasma.  We work in color U(1).
$\overline q\ q$ pair formation
is evaluated according to the Schwinger mechanism using the results of
Brink and Pavel.  The motion of the quarks toward the tube endcaps is
calculated by a Boltzmann equation including collisions.  The tube
undergoes damped radial oscillations
until the electric field settles down to
zero.  The electric field stabilizes the tube against pinch instabilities;
when the field vanishes, the tube disintegrates into mesons.  There is only
one free parameter in the problem, namely the initial flux tube radius, to
which the results are very sensitive.  Among various quantities calculated
is the mean energy of the emitted pions.
\end{abstract}
\newpage

\section{INTRODUCTION}

      Flux tubes are one of the most elementary systems of quantum
chromodynamics.  They are the idealized configurations of heavy
quark-antiquark
pairs at large separation $L$ such that the region between can be assumed to
possess axial-cylindrical symmetry.
They play a central role in lattice QCD calculations and
in models of QCD, as well as in the phenomenology of QCD processes, such as
jets produced in heavy ion collisions or the ``spaghetti" connecting the
separating heavy ions.

      From the
spectroscopy of heavy quarkonia and Regge trajectories,
it has been possible to extract
a reliable measurement of the string tension by assuming the two body
potential for heavy quarks to be of the form
\begin{equation}
V(L)=-{4\alpha_s\over 3 L}  +  \theta L + \ {\rm constant}
\end{equation}
The ``experimental" value is\cite{Casher}
$\theta=913$ MeV/fm = 0.18 GeV$^2$=4.63/fm$^2$.
In fact the linear region of the flux tube potential is not directly
explored by quarkonia or by lattice QCD calculations:

      The wave functions
in quarkonia calculations tend to be centered near the ``knee" of the
potential although high-lying states do explore more deeply into the linear
region.  However other forms of the heavy quark potential (including a
logarithmic term) have also
achieved some success in reproducing quarkonia
spectra.

      Lattice QCD calculations on flux tubes are generally limited to the
quenched approximation (no massless quarks) and allow for a separation of
the heavy quark-antiquark of only about 1 fm, which do not reproduce a
cylindrical region

      Static flux tubes are unstable at separations greater than 1 fm,
since the energy required to stretch the tube by 1 fm is nearly 1 GeV,
which is about the energy difference between a quarkonium, $Q\overline Q$, and
a pair of heavy-light mesons, $Q\ \overline q\ +\ \overline Q\ q$.  Lattice
calculations without light quarks cannot explore this instability.

      Recently there have been several papers~
\cite{Brink,Greiner,Warke,Flintoft} which have explored the
creation of light quark-antiquark pairs as a
mechanism for flux tube breaking.  They
employ a variation of the Schwinger\cite{Schwinger}
parallel plate capacitor model in which
the infinite transverse geometry is replaced by the MIT boundary condition
\begin{equation}
-i\bbox{\gamma\!\cdot\!\rho}\psi(R)=\psi(R)
\end{equation}
at the cylindrical tube
surface for the light quarks.  The electric field ${\bf E}$
is constant over all space in the
longitudinal ($z$) direction, with ${\bf D}$ equal to
${\bf E}$ inside the tube and zero
outside.  It is found that pair production in the
cylinder is suppressed relative to the Schwinger formula due to transverse
confinement which gives the quarks an effective mass proportional to $1/a$
where $a$ is the radius of the flux tube.

      In the present work, we make some simplifying assumptions, the most
significant are:

      We consider infinite flux tubes.  This means that the tube
is long compared with its radius.  For a jet, this means that the
characteristic tube lifetime, $\tau>>a/c$, since the quark producing the
jet travels near the speed of light.

      We use the MIT bag model for the tubes.

      Although some contact is made with color SU(3), we perform our
calculations in U(1).  The extension to Abelian SU(3) is not a serious
complication.

      We adopt one dimensional Debye screening for the quark-quark
interactions used in the Boltzmann collision term.  This is further
simplified to an effective $\delta$-function potential.

      With these approximations, our results depend on only one parameter,
the initial (static) flux tube radius.

     \section{THE MIT MODEL}

      We utilize the framework of the MIT model because it is so very simple.
It is the prototype of color dielectric models.  Some features of various
models of flux tube are discussed in Sec. \ref{sec:mod}.

 In the MIT model, the calculation of the flux tube configuration is
straightforward since it involves no quarks.  The flux through the tube is
equal to the color charge $Q={\textstyle{1\over 2}}\bbox{\lambda} g_s$,
where the
$\bbox{\lambda}$ are the Gell-mann color matrices.
The electric field is $E=Q/A$. Here $Q$ is the charge at one end (say, the
left) of the
flux tube ($\bar Q$ at the other).  It is this charge which will be shielded as
a function of time due to the uniform production of pairs out of the sea
in the flux tube.  We distinguish
it from the charge $q$ of the light quarks being created in pairs,
although for the static model $Q=q$.

The energy per unit length is
given by
\begin{equation}
{{\cal E}\over L} =B\,A + {1\over 2} E^2 \,A
=B\,A + {Q^2\over 2 A}\,
\end{equation}
where $A=\pi a^2$, with $a$ the cylinder radius, and $B$ is the bag
constant.
Minimization with respect to $A$ yields $A=Q/\sqrt{2B}$ and
\begin{equation}
{{\cal E}\over L}\bigg|_{min}\equiv
 \theta=\left[2\,Q^2\,B\right]^{1/2}\,=\,2\,B\,A\,=\,  {Q^2\over A}\,.
\end{equation}

The translation
to QCD obtains by the replacement of $Q^2$ by its expectation value in a
color-singlet state,
\begin{equation}
Q^2\rightarrow=<Q^2>={g_s^2\over 4 } <\bbox{\lambda}^2>
={16\,\pi\alpha_s\over 3 },
\end{equation}
where we have used $<\bbox{\lambda}^2> =16/3$ and $\alpha_s=g_s^2/4\pi.$
\section{THE SCHWINGER MECHANISM}

The rate of pair creation in the Schwinger mechanism with MIT
boundary conditions has been calculated by several groups~
\cite{Brink},\cite{Greiner},\cite{Warke}.
The point of departure  is the QED Schwinger problem of infinite parallel
plate capacitors infinitely far apart\cite{Schwinger}.
The rate of pair production per
unit time per unit volume is
$$
w(\infty)={\kappa^2\over 4 \pi^3}\sum_{n=1}^\infty {e^{-\pi n m_0^2/\kappa}
\over n^2}\qquad\qquad$$
\begin{equation}
\qquad\qquad\rightarrow {\kappa^2\over 4\pi^3}\ {\pi^2\over 6}=
{\kappa^2\over 24\pi}
\qquad {\rm for} \quad m_0\rightarrow 0.
\end{equation}
where $m_0$ is the Fermion mass and $\kappa=|q\,{\bf E}|$.

In the MIT model, $\kappa$ (with $q=Q=\sqrt{\theta\pi}a$) is just the string
tension $\theta$.
in which case
\begin{equation}
w(\infty)\approx 0.25\;c/{\rm fm}^4
\end{equation}
This identification is model-dependent, as discussed below.  For example,
Pavel and Brink\cite{Brink} identify $\kappa=2\theta$.
(For reference, we note that those authors denote the string tension
by $\sigma$ whereas we use $\theta$.)

As we will see below, the rate is strongly suppressed by confinement.

\section{CONFINEMENT AND TRANSVERSE MASS}

Transverse confinement has the effect of introducing a ``transverse mass"
in the formula for $w$: $m_0^2\rightarrow m_0^2+x_n^2/a^2$.
Using the results of Pavel and Brink\cite{Brink} [{\it cf.}
their Eq. (4.3)],
the pair production rate per unit time per unit length $W$ is
\begin{equation}
W(a) =-{n_f\kappa\over \pi}\sum_{x_n>0}\ln\left[1-e^{-\pi x_n^2/a^2\kappa}
	\right]
\approx {n_f\kappa \over \pi} e^{-\pi\, x_1^2/a^2\kappa}\,,
\end{equation}
where $n_f$ is the number of flavors.
For
reasonable $a$ and $m_0=0$, the two lowest
transverse modes, $x_{\pm 1}$ dominate, and we consider these only
here: $|x_{\pm 1}|=1.4347$.  Henceforth we will assume $m_0=0$.

\section{SIMPLE ESTIMATE OF COLLAPSE TIME}

In the following, we work in QED, which is to say, charge is a scalar
quantity, plus or minus.

We now consider $Q=Q(t)$, since the charge at the ends is shielded due to the
flow of opposite charge into the ends.  We will calculate it more carefully
in the next section, but it is a good approximation to calculate the
current by using the initial radius of the flux tube and the electric
field for short times.  The number of pairs is proportional to the time,
and the mean velocity of a quark is proportional to the time.  It is then
easy to show that
\begin{equation}
{dQ(t)\over dt}=-2\,J\approx
-2\,W(0)\,q{<p>\over m}\,t\approx
-2\;{W(0)\,q^2\over 2\,A\,m}\,t^2\,Q(0)\,,
\end{equation}
where $J$ is the current
carried by positive quarks; negative quarks carry an equal current,
including the same sign.  Hence the factor of 2.
We have used the average velocity of quarks $<p>/m\approx
Q(0)\,q\,t/(2\,A\,m)$,
We make the identification $m=m_\perp= x_1/a$.  Again we note that
$a$ is the tube radius at $t=0$.
One more time integration gives
\begin{equation}
Q(t)=Q(0)\,\left[1-{W\,q^2\over 3\,A\,m}\,t^3\right]\,.
\end{equation}
We now estimate the characteristic lifetime of the tube to be
\begin{equation}
\tau(a)\approx \left[{3\,m\,A\over W\,q^2}\right]^{1/3}
  =\left[{3\,m\over W\,\kappa}\right]^{1/3}
  =\left[{3\, \pi\,x\,e^{\pi\, x_1^2/a^2\kappa}\over \kappa^2
\,a}\right]^{1/3}\,.
\end{equation}
Note that there is no dependence on the length of the flux tube!

Taking $n_f=2$, we find
\begin{eqnarray}
   \kappa&=&\ \ \theta\qquad\,\ 2\theta \\
\tau(0.5)&=&7.0\qquad 1.7\,\ \quad {\rm fm}/c\,, \\
\tau(1.0)&=&1.4\qquad 0.68\quad {\rm fm}/c\,.
\end{eqnarray}
Recall that $\kappa=\theta$ in the MIT model.

\section{COLLISIONLESS BOLTZMANN EQUATION}

      Here we still employ the adiabatic MIT model to follow the dynamics
of the flux tube wall by assuming that the tube radius is in instantaneous
equilibrium.
In addition to the electric field, the light quark plasma also exerts a
pressure on the tube wall.  We assume that only the lowest transverse
eigenmode is excited ($x_{\pm 1}$), so that the tube radius, now denoted by
$R(t)$ instead of $a$,
is determined
by minimizing the transverse energy
\begin{equation}
{{\cal E}\over L}=\pi\,B\,R^2+{Q^2\over 2\pi\,R^2}+{2\,N\, x_1\over R},
\end{equation}
which yields
\begin{equation}
-{Q^2\over \pi}+2\pi\,B R^4-2\,N\,x_1R=0.
\end{equation}
$N=N(t)$ is the number of {\it pairs} per unit length,
\begin{equation}
N(t)=\int_{-\infty}^{\infty} n(p,t)\,dp=\int_0^t W(t')\,dt \label{eq:N}
\end{equation}
with $n(p,t)$
the momentum density per unit length for either $q$ or  $\overline q$ .
Note that we distinguish between the shielded heavy quark charge $Q(t)$ and
the elementary light quark charge $q$, where $Q(0)=q$.

The pair production rate (per unit length) is given by
\begin{equation}
W=n_f\,{\kappa\over \pi}e^{-\pi x_1^2/R^2\kappa}\,,
\end{equation}
with $\kappa=q|E|$.

The transverse motion of the quarks is now treated relativistically but
classically.  Thus the relativistic mass is taken to be
\begin{equation}
m^2(t)=\left({x_1\over R(t)}\right)^2+p^2\,.
\end{equation}
      The rate of change of $n(p,t)$ is governed by the
coupled integro-differential equations:
\begin{equation}
{\partial\,n\over \partial t}=-q\,E\,{\partial n \over \partial p}+
      \delta(p)\,W + \left({dn\over dt}\right)_{col}\,,  \label{eq:bol}
\end{equation}
\begin{equation}
J(t)=q\int dp\,n\,p/m\,,
\end{equation}
\begin{equation}
{d\,Q\over dt}=-2\,J\,,
\end{equation}
\begin{equation}
 E(t)={Q(t)\over\pi\,R(t)^2}\,,
\end{equation}
where $(dn/dt)_{col}$ is the rate of change of the momentum distribution due
to collisions.
The initial conditions are $n(p,0)=0,\quad Q(0)=q=\sqrt{\theta\pi}a,
\qquad R(0)=a.$

We wish to point out the very interesting fact that {\bf no magnetic
fields} are produced!  This is because, in Maxwell's famous fourth
equation, the convection current is exactly cancelled by the displacement
current.

One valuable check on the numerics is provided by the agreement between the
two forms of Eq. \ref{eq:N}.  Excellent agreement is attained
until numerical instability sets in, when the two values diverge
dramatically, in which case time or momentum steps were changed until
equality was achieved.

These equations have been solved numerically first without the collision
term for several parameter sets.
The results are displayed in figure \ref{fig2};  the thin lines are for the
collisionless solutions.  We note here oscillations  with negligible
damping.  In the absence of damping, the current continues to flow even
when the charge at the end caps (and hence ${\bf E}$ goes through zero.
Consequently the charge at the end caps changes sign, tending to reverse
the current. The number of pairs increases with time.  The energy
per unit length also increases; there is no energy conservation within a
segment of the tube.

\section{COLLISIONAL THERMALIZATION AND DAMPING}


The problem is simplified by the one-dimensional nature of the dynamics: we
assume that the transverse motion can be handled by the adiabatic MIT
model.  Scattering between quarks (charge $q$) or between quarks and
anti-quarks (charge $-q$) is like that of
beads on a string.  The colliding particles exchange momentum with some
probability to be computed below.  Collisions between identical particles
does not alter the momentum distribution -- it is as though no collision
took place.  Since the interaction is independent of spin and flavor, we do
not consider any quark-quark collisions, but only
collisions between quarks and antiquarks.
Then the change in population density of quarks in momentum satisfies
the equation
\begin{eqnarray}
\left({dn_+(p_1)\over dt}\right)_{col}&=
      &\int dp_2\,|{\cal R}(p_1-p_2)|^2 |(v_1-v_2|[n_-(p_1)n_+(p_2)-
                                        n_+(p_1)n_-(p_2)] \nonumber\\
&=&\int dp_2\,|{\cal R}(p_1-p_2)|^2 |(v_1-v_2|[n(-p_1)n(p_2)-n(p_1)n(-p_2)]
\end{eqnarray}
with $n=n_+$.  Here $|{\cal R}|^2$ is the reflection probability for the
collision of quarks with antiquarks in a one dimensional potential barrier
or well.  The potential should take into account shielding introduced by
the presence of other quarks and antiquarks.

\subsection{The Shielded Quark-Antiquark Interaction}

The interaction between confined quarks
is that of disks in the tube of radius $R$.  The confinement of the
$D$-field
by the tube walls leads to a potential between bare quarks
which is linear in the separation,
just like the flux tube potential itself.  However the quarks are shielded
by other quarks and antiquarks.  An estimate of the shielding
can be obtained from a one-dimensional
version of the Debye formula:

The charge density in the vicinity of a quark at (say) $z=0$
in a thermal bath in the rest frame of the quark is
\begin{equation}
\rho(z)={q\over \pi\,R^2}\delta(z)
+C\int_{-\infty}^\infty dp\left[e^{\epsilon^+(p)/k_BT}-e^{\epsilon^-(p)/k_BT}
      \right]
\label{eq:rho}				\end{equation}
where
\begin{equation}
\epsilon^{\pm}(p)=\pm q\phi(z)+\sqrt{m_0^2+p^2}
\label{eq:eps}				\end{equation}
with $m_0=x_1/R$ and $\phi(z)$ the shielded potential  to br determined.
Since $\phi(z=\pm\infty)=0$, it follows that the polarization charge of
+/$-$
quarks is
\begin{equation}
\rho^{\pm}(x=\pm\infty)=\pm{ q\,N \over \pi R^2}=\pm C\int_{-\infty}^\infty
      e^{-\sqrt{m_0^2+p^2}/k_B T}dp
\end{equation}
and hence
\begin{equation}
\rho(z)={q\over \pi R^2}\delta(z)+{q\,N\over \pi R^2}\
   \left[e^{-q\phi(z)/k_BT}-e^{q\phi(z)/k_B}      \right]
\end{equation}

If, as turns out to be the case generally, $|q\phi)|/k_BT<<1$, the
expression can be linearized in $\phi$, and we obtain
\begin{equation}
\rho(z)={q\over \pi R^2}\delta(z)-{2N\,q^2\over\pi R^2k_BT}\phi(z)
\end{equation}
Poisson's equation in one dimension then reads
\begin{equation}
{d^2\phi\over dz^2}=-\rho(z)=-{q\over \pi R^2}\delta(z)+{2N\,q^2\over
      \pi R^2k_BT}\phi(z)
\end{equation}
The solution to this equation is given by
\begin{equation}
\phi(z)={q\lambda_D\over 2\pi R^2}e^{-|z|/\lambda_D}
\end{equation}
with the Debye length given by
\begin{equation}
\lambda_D^2={\pi R^2k_BT\over 2\,N\,q^2}
\end{equation}
Therefore the quark-antiquark potential is given by
\begin{equation}
V_D(z)=-q \phi(z)=-{q^2\lambda_D\over 2\pi R^2}e^{-|z|/\lambda_D}
\label{eq:VD}  \end{equation}
\subsection{Calculation of the reflection probability}

It is an elementary quantum mechanical problem to calculate $|{\cal R}|^2$
for the one dimensional Debye potential  Eq. \ref{eq:VD}.
For greater simplicity and
insight, we consider two analytic approximations for $|{\cal R}|^2$:

(1) A square-well potential of width $D$ and depth $V_0$ such that $V_0D$
equals the integral of $V_D(z)$. A reasonable value for $D$ is
$2\lambda_D$.  The reflection probability can be found in any of
several texts.

(2) A delta-function potential of strength $V_\delta$ (like the square-well
potential with $D\to 0$).  The strength is given by

\begin{equation}
V_\delta=-{k_BT\over 2\,N}
\end{equation}

\subsection{Evaluating the temperature}
In order to identify the temperature $T$, we note that, except for very
small times, the charactersitic momenta in the problem are large compared
with $m$, and therefore we may set $\epsilon^\pm\approx \pm q\phi+|p|$ in
Eq. \ref{eq:eps}.  Consequently we can approximate [cf. Eq. \ref{eq:rho}]
\begin{equation}
n(p)\equiv n_+(p)\approx Ce^{-|p-<p>|/k_B T},
\end{equation}
where $<p>$ is the average drift momentum.  For this form of the Boltzmann
distribution, we have
\begin{equation}
N=\int_{-\infty}^\infty n(p)\,dp=2\, C\,T
\end{equation}
and the fluctuation in the momentum yields the temperature:
\begin{equation}
(k_B T)^2=(2N)^{-1}\int_{-\infty}^\infty n(p) [p-<p>]^2dp
\end{equation}

In the calculation, we use these equations to determine the temperature
at each time from the calculated $n(p,t)$.  Then
\begin{equation}
(k_BT)^2=<(p-<p>)^2>/2
\end{equation}
where $<\ >$ means momentum average with respect to the calculated $n(p,t)$.

\subsection{${\cal R}$ for the delta-function}

The delta-function case $V(z_1-z_2)=V_\delta \delta(z_1-z_2)$ has a
particularly simple form for the reflection amplitude ${\cal R}$ and
probability $|{\cal R}|^2$ for particles of momenta $p_1$ and $p_2$, namely
\begin{equation}
|{\cal R}|^2={V_\delta^2\over V_\delta^2 + (p_1-p_2)^2/\mu^2}\label{eq:vd}
\end{equation}
where, without rigor, we have used  a reduced mass
\begin{equation}
{1\over \mu}={1\over m(p_1)} + {1\over m(p_2)}\,.	\label{eq:mu}
\end{equation}
With the self-consistent determination of $V_\delta$ from Eqs. \ref{eq:vd}
and \ref{eq:mu}, a complete solution of Eq. \ref{eq:bol} and comparison
with the collisionless results for the same model parameter (only $a$) is
possible.

\subsection{Damping time}

We see here the evolution from a {\it flux} tube to a {\it plasma} tube.
The behaviour of this evolution depends strongly on the single parameter
$a$.  For $a$ between 0.5 and 1.5 fm, we note damped oscillations which
increase in amplitude and decreases in period as $a$ increases.  We proffer
three measure of the characteristic damping time, $\tau_0,\ \tau_1,$ and
$\tau_2$.  $\tau_0$ is defined as the time at which $Q(t)$ first crosses
zero.  The other definitions are
\begin{equation}
\tau_1\equiv \int_0^\infty |Q(t)|\,t\,dt\bigg/_0^\infty |Q(t)|\,dt
\label{eq:t1}
\end{equation}
and
\begin{equation}
\tau_2\equiv \int_0^\infty Q^2(t)\,t\,dt\bigg/\int_0^\infty Q(t)^2\,dt
\label{eq:t2}
\end{equation}
Calculated results for these three measures are shown in Fig. \ref{fig6}.

In Fig. \ref{fig5} we show the number of pairs produced
per unit length, and the
energy per pion. The pion production is calculated
assuming that one meson, $\pi,\ \rho,$ or $\omega$, is produced in
proportion to their spin-isospin statistical weights: 3/15 for pions, 9/15
for rhos, and 3/15 for omegas.  The rho and omega decay to 2 and three
pions respectively.  The total energy per unit length is distributed
among the
pions per unit length.  The result is shown in Fig. \ref{fig6}.

      Note that our calculations do not depend upon the motion of the end
caps, and therefore, within our approximations, the energy per ion is flat
in rapidity.

      Our results show a high degree of sensitivity to the initial flux
tube radius $a$.  This provides a useful handle for distinguishing among
various flux tube models and, in particular, the appropriate size of $a$.

\subsection{Ultimate disintegration of the plasma tube}

      Unlike the plasmas in a fusion reactor,
the {\it flux} tube is stable
against the sausage instability; the difference is due to the absence of
magnetic fields.  After the electric field had decayed, this stabilizing
effect is removed and the {\it plasma} tube falls apart.  We have not
attempted yet to follow this final phase, but use the various $\tau$'s
given above to estimate the disintegration time.  Our own bias is to use
$\tau_1$.

\section{MODELS OF FLUX TUBES}
\label{sec:mod}

Since the collapse time is strongly dependent upon the initial flux tube
radius $R(0)=a$,
we now address more realistic flux tube calculations to determine $a$ and
the relationship between $\kappa$ and $\theta$.

\subsection{The MIT Model}

In this model, the field energy $E^2/2$ is uniform within a radius $a$
which is characteristically 1 fm or larger.  The flux tube energy is shared
equally between the electric field and volume energies:
$\theta=E^2A=Q^2/A=q|E|$.

\subsection{Lattice Gauge Calculations}

Lattice calculations at present are usually restricted to the quenched
approximation (no zero-mass dynamical quarks) and to $Q\ \overline Q$
separations $L$ of less than about 1 fm.  Nevertheless Sommer\cite{Sommer}
obtains a
dependence of the (longitudinal) electric field energy which appears to be
stable with $L$.  The functional form is roughly exponential, $E^2/2
\propto
 e^{-r_{\perp}^2/b}$ with $<r_{\perp}^2>=6\, b^2\approx (0.2\ {\rm fm})^2$.
The {\it equivalent} radius (the radius of a square form giving the same
$<r_{\perp}^2>$) is $r_{eq}=\sqrt{2}<r_{\perp}^2>^{1/2}\approx 0.3 $ fm.
Here $\theta=2 q\,|E|$.

\subsection{Dual Superconductivity Model}

The dual superconductivity model \cite{Baker} yields a somewhat larger
radius than lattice calculations, with
$r_{eq}=\approx 0.4 $ fm.

\subsection{The CDM}

The chromodielectric model\cite{Fai} is similar to the MIT model,
except that the
electric field energy has a diffuse surface and the flux tube energy $\theta$
also contains a surface term such that $\theta>q|E|$.

\section{FURTHER CORRECTIONS}

There are several effects which are yet to be included in order to improve
the estimate of the collapse time:

(1)  Finite length effects also increase the collapse time.  There are at
least two such effects to consider: a) The current associated with a given
momentum group terminates when all the charge in the group is transported
the full length of the tube. b) No pairs can be created unless the change
in potential the length of the tube is greater than $m_\perp$:
$|q\,E\,L|>x_1/R$.  These also lengthens the collapse time.

(2)  Dynamics of the confinement mechanism and associated energy
conservation.  One might consider the chromodielectric soliton model to
handle this.

\section{FINAL COMMENTS}

It is important to note that the calculations presented here do not depend
on the velocity of the heavy quarks in the longitudinal direction.
The collapse time depends sensitively on the radius of the flux tube and
indeed on detailed structure of the tube.

For different approaches to the problem, see Kluger, {\it et
al.}\cite{Kluger} and Rau\cite{rau}.

\acknowledgments

We wish to thanks Profs. M. Baker, H. A. Pirner and R. Sommer for useful
discussions.  One of us (LW) wishes to thank Prof. M. Birse for for many
enjoyable discussions about flux tubes.
This work is supported in part by the U. S. Department of
Energy.

\begin{figure}
\noindent \caption {Schematic evolution of a flux tube to a plasma
tube to a
collection of mesons.
The radius of the tube exhibits damped oscillations.
\label{fig1}}

\vskip 20pt

\noindent \caption{Variaton with time of the flux tube radius
$R(t)/a$ (upper curves)
and net endcap charge $Q(t)/q$ (lower curves) for various $a$.  For the
case (a) of $a=0.5$ fm. we have also shown in thin lines the solution to
the
collisionless Botzmann equation. {\hskip 15cm}\label{fig2}}
\medskip

\noindent \caption{The mean longitudinal momentum per quark as a
function of time for $a=0.5$ and 1.0 fm. {\hskip 30cm}
\label{fig3}}
\medskip

\noindent \caption{The quark ``temperature" as a function of time for $a=0.5$
and 1.0 fm.
{\hskip 10cm}
\label{fig4}}
\medskip

\noindent \caption{The growth of the number of pairs as a function of time
for $a=0.5$. {\hskip 5cm}
\label{fig5}}
\medskip

\noindent \caption{The variation of several important quantities with $a$:
The characteristic damping times (in fm/c)
$\tau_0$ (where $Q(t)$ first crosses
zero), $\tau_1$ and $\tau_2$ (as defined by equations \ref{eq:t1} and
\ref{eq:t2}).  Ten times the number of pairs produced per fm is
denoted by 10 N.  $E$ is the final energy per fm (1/fm$^2$).
$E_\pi$ is the
mean energy (in GeV) of pions produced by disintegration of the tube.
\label{fig6}}
\end{figure}

\end{document}